\documentclass[prl,twocolumn,amsmath,amsfonts,amssymb]{revtex4} 
\usepackage{epsfig,psfrag,graphicx,bm,color}
\begin{document}
\title{Gamma-ray and Radio Constraints of High Positron Rate Dark Matter Models Annihilating into New Light Particles}

\author{Lars Bergstr\"om$^a$}
\author{Gianfranco Bertone$^b$}
\author{Torsten Bringmann$^a$}
\author{Joakim Edsj\"o$^a$}
\author{Marco Taoso$^{b,c}$}
\affiliation{$^a$Oskar Klein Centre for Cosmoparticle Physics, Department of Physics, Stockholm University, AlbaNova, SE - 106 91 Stockholm, Sweden}
\affiliation{$^b$Institut d'Astrophysique de Paris, UMR7095-CNRS \\ Universit\'e Pierre et Marie Curie, 98bis Boulevard Arago, 75014 Paris, France}
\affiliation{$^c$ INFN, Sezione di Padova, via Marzolo 8, Padova, 35131, Italy}



\begin{abstract}
The possibility of explaining
the positron and electron excess recently found by the PAMELA and ATIC collaborations in terms of dark matter (DM) annihilation
has attracted considerable attention.
Models surviving bounds from, e.g, antiproton
production generally fall
into two classes, where either DM annihilates directly with a large
branching fraction into light leptons, or, as in the recent models of
Arkani-Hamed et al., and of Nomura and Thaler, the annihilation gives
low-mass (pseudo)scalars or vectors $\phi$ which then decay into $\mu^+\mu^-$ or $e^+e^-$.
While the constraints on the first kind of models have recently been treated by several authors, we study here specifically  models of the second type
which rely on an efficient Sommerfeld enhancement in order to obtain the necessary boost in the annihilation cross section.
We compute the photon flux generated by QED radiative corrections to the decay of $\phi$
and show that this indeed gives
a rather spectacular broad peak in $E^2d\sigma/dE$, which for these extreme
values of the cross section violates gamma-ray observations of the Galactic
center for DM density profiles steeper than that of Navarro, Frenk and White.
The most stringent constraint comes from the comparison of the predicted synchrotron
radiation in the central part of the Galaxy with radio observations of
Sgr A*. For the most commonly adopted DM profiles, the models
that provide a good fit to the PAMELA and ATIC data are ruled out, unless
there are physical processes that boost the local anti-matter fluxes
more than one order of magnitude, while not affecting the gamma-ray or radio fluxes.

\end{abstract}

\maketitle

\newcommand{\ga}{\gamma}
\newcommand{\be}{\begin{equation}}
\newcommand{\ee}{\end{equation}}
\newcommand{\bea}{\begin{eqnarray}}
\newcommand{\eea}{\end{eqnarray}}
\newcommand{\code}[1]{{\tt #1}}

\hyphenation{} There have recently been indications of a very
interesting enhancement in the amount of cosmic ray electrons and
positrons detected near the Earth, both seen by PAMELA in the ratio
of positrons to the sum of electrons and positrons between a few GeV
and 100 GeV \cite{pamela}, and by ATIC in the sum of electrons and
positrons at several hundred GeV to 1 TeV \cite{atic}. While these
so far unexplained excesses might be due to standard astrophysical
processes \cite{astrophysical}, positrons also constitute one of the
promising channels in which to search for dark matter (DM; for reviews,
see \cite{reviews}),  and these new experimental findings have
therefore already triggered  a large number of theoretical analyses
trying to explain the data as being induced by DM
annihilation or decay (see e.g. Ref.~\cite{BBE08} and references
therein for supersymmetric DM, Refs.~\cite{post-pamela} for
alternative DM scenarios and Refs.~\cite{decaying} for decaying
DM scenarios). In general, these analyses seem
to point at the need for DM particles with masses in the TeV range
that annihilate, with a very large rate, dominantly into charged
light leptons.

The bremsstrahlung process, falling like $E_\gamma^{-1}$, is generally regarded in particle physics as having
a ``soft'' spectrum. In the astrophysical context,
this is, however, on the contrary a quite hard spectrum, since most of the background $\gamma$-ray
spectra like those from acceleration near supernova remnants usually fall like $E_\gamma^{-2}$ or faster.
Gamma-rays from DM generally feature a spectrum that is somewhere in between these two
at low energies ($E_\gamma^{-1.5}$) and drops even faster close to the DM particle
mass \cite{BUB} (for important exceptions see, however,
\cite{BBE}.) If the DM particles $\chi$ annihilate directly into a pair of
charged leptons, the photon distribution from the process $\chi\chi\rightarrow\ell^+\ell^-\gamma$, for $m_\chi\gg m_\ell$,
is to a good approximation of the Weizs\"acker-Williams form (see, e.g., \cite{FSR}):
\bea
\label{dNdx}
 &&\frac{d(\sigma v)}{dx} = (\sigma v)_{\ell\ell}{\alpha_\mathrm{em} \over\pi}
{((1-x)^2+1)\over x}\ln \left[{4m_\chi^2(1-x)\over m_\ell^2}\right]\,,
\label{eq:brems}
\eea
where $x=E_\gamma/m_\chi$ and $(\sigma v)_{\ell\ell}$ is the annihilation rate for the lowest order
process $\chi\chi\rightarrow\ell^+\ell^-$ (Note that the above approximation also breaks down when
there is a symmetry that suppresses the annihilation into two-body, but not into three-body final states \cite{bergstrom89}).

This case has recently been treated by
\cite{Bell:2008vx,Bertone:2008xr,Cholis:2008wq}. (The last of these references also briefly treats, but leaves for a more detailed calculation, the kind of processes we will compute here.)
It was found
that the gamma-rays produced in DM models with these annihilation
modes lead to rather severe constraints. Even more stringent bounds
on this type of DM models that try to explain the PAMELA and ATIC
data arise from the synchrotron radiation produced by the resulting
population of electrons and positrons, in realistic models of the
DM density distribution and for a wide variety of
assumptions about the magnetic field in the inner Galaxy
\cite{Regis:2008ij,Bertone:2008xr}.

It remains to consider another possibility, where  DM annihilates into a new type of light (sub-GeV) particles $\phi$
that in turn dominantly decay into light leptons (see \cite{arkani-hamed} for a general account of this idea).
The advantage of this type of models is that the strongly constrained decay into hadronic modes (see, e.g., \cite{Cirelli:2008pk})
is kinematically forbidden and that  Sommerfeld enhancements in the limit of the small galactic DM velocities expected today allow for
the very large annihilation cross sections that are needed to explain the PAMELA/ATIC results, but which at first seem to be at
odds with the cross sections required to get the right thermal relic density for the DM.
Another interesting feature of the Arkani-Hamed et al.\ model \cite{arkani-hamed} is that it encompasses ideas that have been proposed to explain the WMAP haze \cite{wmap-haze} and the INTEGRAL excess \cite{Finkbeiner:2007kk}.
\begin{table}[t]
\begin{tabular}{|c|c|c|c|c||c|c|c|}
\hline
\multicolumn{5}{|c||}{Arkani-Hamed et al. type} & \multicolumn{3}{|c|}{Nomura-Thaler type}\\
& $m_\phi$[GeV] & type & $e^+e^-$ & $\mu^+\mu^-$ & & $m_s$[GeV]& $m_a$[GeV]\\
\hline
AH1 & $0.1$ & scalar & 100\% & -    & $N1$ & $5$  & $0.5$ \\
AH2 & $0.1$ & vector & 100\% & -    & $N2$ & $20$ & $0.36$ \\
AH3 & $0.25$ & vector & 67\% & 33\% & $N3$ & $20$ & $0.5$  \\
AH4 & $0.25$ & scalar & - & 100\%   & $N4$ & $20$ & $0.8$  \\
     &        &   &   &         & $N5$ & $50$ & $0.5$  \\
\hline
\end{tabular}
\caption{Our benchmark scenarios.
}
\label{tab:bm}
\end{table}
As pointed out in \cite{arkani-hamed,Cholis:2008qq}, one may basically distinguish between scalar and vector $\phi$ and
whether or not $m_\phi\lesssim2m_\mu$ (in which case it dominantly decays into $e^+e^-$). For $m_\phi\gtrsim m_\pi$,
even decays into pions should be taken into account (which we neglect here). While
$m_\phi\gtrsim10\,$MeV is roughly needed not to be in conflict with Big Bang Nuclesynthesis, one has to require $m_\phi\gtrsim100\,$MeV
in order to get Sommerfeld enhancements of the order $10^3-10^4$ that are needed to explain the PAMELA/ATIC result with these types of DM models.
Based on this discussion, we adopt the four benchmark settings A1--A4 summarized in Tab.~\ref{tab:bm}.

While \cite{arkani-hamed} describes a rather general set-up,  \cite{nomura} introduces a concrete realization of this idea; the proposed model
has the appealing feature of
containing a ``standard'' Peccei-Quinn axion and can  be embedded
in a fully realistic supersymmetric scenario.
Here, DM annihilates into a scalar $s$ and
a pseudoscalar $a$, $\chi \chi \rightarrow s a$. With a mass scale of $360\,\mathrm{MeV}\lesssim m_a\lesssim800\,\mathrm{MeV}$, the latter mostly decays into muons,
which subsequently
decay into electrons or positrons. The benchmark models for this setup N1--N5 are also given in Tab.~\ref{tab:bm}.

For the first $a$ particle created in the $\chi\chi$ annihilation, we analytically compute the photon
multiplicity $(dN/dE_\gamma)^{(a)}$ from $a\rightarrow\mu^+\mu^-\gamma$  in the rest frame of $a$.
We then make a Lorentz boost back to the DM frame, i.e.\ the Galactic rest frame, to get
\be
\label{eq:boost}
\left(\frac{dN}{dE_\gamma}\right)^{(\mathrm{DM})}=\frac{1}{2\beta\gamma}\int_{E/(\gamma(1+\beta))}^{E/(\gamma(1-\beta))}\frac{dE'}{E'}\left(\frac{dN}{dE'_\gamma}\right)^{(a)}\,,
\ee
with
$\gamma=(m_\chi/m_a)\left[1-(m_s^2-m_a^2)/(4m_\chi^2)\right]$
since the annihilation takes place essentially
at rest (typical galactic velocities are $10^{-3}$). Axions resulting from
$s\rightarrow aa$ we treat in a similar way, boosting them first to the $s$-frame and from this to the DM frame.
Since $s$ may have a mass up to 50 GeV, the gamma-ray spectrum may even receive important contributions
from its decay into bottom quarks or tau leptons, a possibility which we will shortly return to.
(Bremsstrahlung from electrons in the muon decay will give $\gamma$s of lower energies and will thus not be important for our constraints.)

\begin{figure}[t!]
\includegraphics[width=\columnwidth]{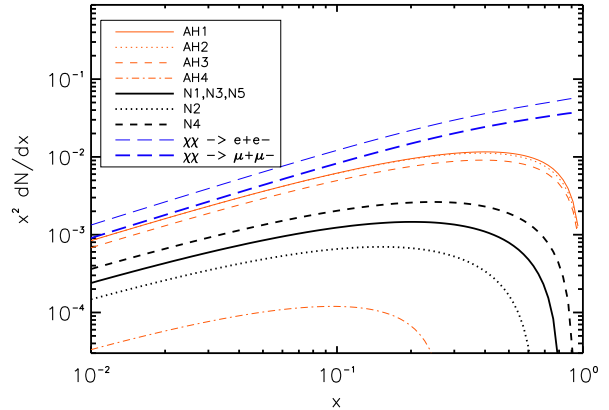}
\caption{The various possible photon spectra that can arise from DM annihilating to new
light particles which in turn decay into charged leptons. For the
 models $N1$ -- $N5$, we neglect here the decay of $s$ to tau-leptons or bottom quarks -- see Fig.~\ref{fig:dnde_nfw} for an example of how this changes the spectra. For comparison, we also indicate the spectrum from DM directly annihilating to charged leptons.
\label{fig:dNdx}}
\end{figure}

Summing up all these contributions, we arrive at the total photon spectrum in the DM frame that we show in Fig.~\ref{fig:dNdx} for the models N1--N5 in Tab.~\ref{tab:bm}. We also include the corresponding spectra obtained in the Arkani-Hamed et al.~set-up (models A1--A4) and, for comparison, the case of 1 TeV DM particles directly annihilating into $e^+e^-$ or $\mu^+\mu^-$. Please note that, from Eq.~(\ref{eq:boost}), the quantity $dN/dx$ for the models listed in Tab.~\ref{tab:bm} is independent of $m_\chi$ as long as $m_\chi\gg m_a,m_s$; the direct annihilation of DM into leptons, on the other hand, \emph{does} contain a logarithmic dependence on $m_\chi$.
 Let us mention that while Eq.~(\ref{dNdx}) provides a rather good approximation to our analytic results for photons radiated from $e^+e^-$ pairs, it overestimates the photon yield from muons (especially when the mass of the decaying particle is close to $m_\mu$ like, e.g., in model AH4).

\begin{figure}[t!]
\includegraphics[width=\columnwidth]{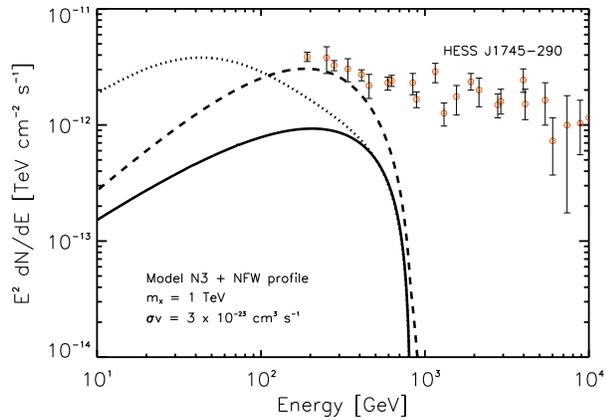}
\caption{The total gamma-ray spectrum $dN/dE_\gamma$, for an NFW halo, from a 1 TeV
DM particle annihilating into a pseudoscalar $a$ (decaying to muons) and a scalar $s$ which decays to $aa$ (solid line) or only in $95\%$ of the cases into $aa$ and in $5\%$ into $b\bar b$ (dotted line) or $\tau^+\tau^+$ (dashed line). The masses for $a$ and $s$ are those of model $N3$ of Tab.~\ref{tab:bm}, so the solid line corresponds to the $N3$-line shown in Fig.~\ref{fig:dNdx}.
\label{fig:dnde_nfw}}
\end{figure}

Once a DM profile $\rho(r)$ is assumed, it is straightforward to estimate
the corresponding gamma-ray flux from a solid angle $\Delta\Omega$ towards the galactic center:
\begin{equation}
\label{gammaflux}
\frac{d \Phi_\gamma}{dE} = \frac{1}{8\pi} \frac{\sigma v}{m_\chi^2} \frac{dN_\gamma}{dE} \int d\lambda \int_{\Delta\Omega} d\Omega \rho^2(\lambda)\,
\end{equation}
where $\lambda$ is the line of sight distance.
In Fig.\;\ref{fig:dnde_nfw}, we compare the resulting flux to the gamma-ray data from the galactic center taken by the H.E.S.S. telescope \cite{hessgc}, which has an angular resolution of about $0.1^\circ$, thus $\Delta\Omega=10^{-5}\,$sr. We here show the spectrum for model N3 and, for comparison, the case where $s$ decays not only to axions but with a branching ratio of 5\% to $\bar b b$ or $\tau^+\tau^-$ (which is the typical case for the model presented in \cite{nomura}). By comparison with Fig.~\ref{fig:dNdx}, it is straightforward to arrive at the corresponding spectra for the other models in Tab.~\ref{tab:bm}. We have here adopted a so-called Navarro-Frenk-White profile \cite{nfw}, with the same
parameters as in Ref.~\cite{Bertone:2008xr}. Note that the gamma-ray spectra in this
case are consistent with the HESS data, unlike the case of the annihilation modes
discussed in  \cite{Bertone:2008xr}, for the same density profile. Assuming a profile
$\rho(r) \propto r^{-1.2}$, as needed to explain the WMAP 'Haze' (see Ref.~\cite{wmap-haze}), the constraints become much more stringent. However, at the
same time they become
much more sensitive to the dependence of $\sigma v$ on the velocity dispersion of
DM, which inevitably increases in the vicinity of the supermassive black hole at
the Galactic center.
As we shall see soon, however, it is possible to derive even tighter
constraints without making assumptions on the small-v behaviour of $\sigma v$.

Before that, however, let us note that another potential source of gamma rays from DM annihilations are dwarf galaxies, like the Sagittarius dwarf galaxy, observed by HESS \cite{hess-sgr}. The HESS observations put an upper bound on the integrated gamma flux above 250 GeV of $\Phi_\gamma < 3.6 \times 10^{-12}$ cm$^{-2}$s$^{-1}$. Assuming an NFW (isothermal) profile in the Sagittarius dwarf galaxy, this can be translated to the limit $\sigma v < 7.4\times10^{-22}$ ($2.2\times10^{-23}$) cm$^{3}$ s$^{-1}$ for model N3. For the other models in Tab.~\ref{tab:bm}, the limits differ by a factor of a few as indicted by the spectra in Fig.~\ref{fig:dNdx}. For other dwarf galaxies, the limits are similar: using a conservative estimate of the line of sight integral from Ref.~\cite{strigari}, the limits on the gamma flux from Willman~1 as observed by Magic \cite{magic-wil1}, e.g., translate to $\sigma v < 1.3 \times 10^{-21}$ cm$^{3}$ s$^{-1}$. However, the uncertainties from dynamical constraints \cite{strigari} are large and improved future data might result in better constraints. As one typically needs a boost of order $10^3$ to explain the PAMELA data, we note that the limits derived here are very close to the required $\sigma v$. This means that for some models, like AH1--AH3, the more optimistic scenarios for the halo profile of e.g.~the Sagittarius dwarf are excluded.

A rather stringent constraint on the rate of injection of high energy
$e^\pm$ in the Galaxy comes from the analysis of the synchrotron radiation
produced by these particles as they propagate in the Galactic magnetic field.
Although observations of different targets and at different wavelengths
provide interesting constraints~\cite{Bertone:2002ms}, the most stringent ones come from
radio observations of the Galactic center, where the DM density
is highest~\cite{Bertone:2002ms,radiogc,Regis:2008ij}.

The synchrotron luminosity generated by a distribution of electrons and positrons
produced by a DM distribution with profile $\rho(r)$ in a magnetic field $B(r)$ is
\begin{equation}
\label{eq:syn}
\nu L\nu = 2\pi \frac{\sigma v}{m_\chi^2} \int dr \,r^2\, \rho^2(r) E_p Y_e(E_p)
\end{equation}
where $E_p=\nu^{1/2}[0.29 (3/4\pi)(e/m_e c^2)^3B(r)]^{-1/2}$, $Y_e(E)=\int_E^{m_\chi} dE' \, dN_e/dE'$
and we have adopted the monochromatic approximation for the synchrotron emission, assuming $P(\nu,E) = (8\pi/9\sqrt{3})\,\delta(\nu / \nu_c-0.29)$,
with $\nu_c=(3eB E^2)/(4\pi m_e^3 c^6)$, for its spectrum.

\begin{figure}[t!]
\includegraphics[width=\columnwidth]{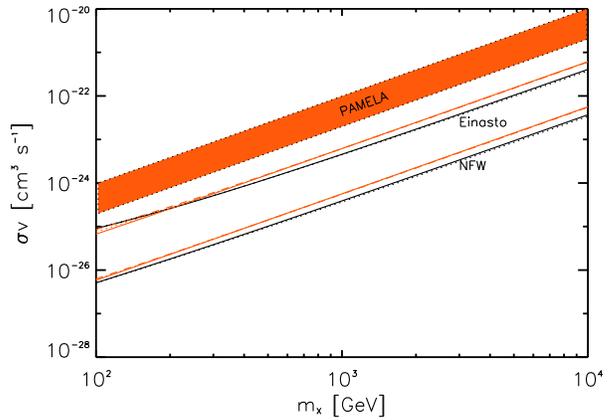}
\caption{Exclusion plot in the $\sigma v$ vs. mass plane.
The two sets of curves give the maximum annihilation cross section compatible
with radio observations of Sgr A* for  Einasto and NFW  profiles.
The color code of the curves is the same as in Fig.~\ref{fig:dNdx}. The shaded region, corresponding
to the range of annihilation cross sections that provide a good fit to the
PAMELA and ATIC data, appears to be in conflict with observations, unless the
DM profile is more shallow than Einasto.}
\label{fig:radio}
\end{figure}

By comparing the predicted synchrotron radiation with radio observations,
we can set limits on the annihilation cross section for any given
annihilation channel, following a procedure similar to Ref.~\cite{Bertone:2008xr}.
The most stringent constraint comes from the upper limit on the
radio emission from a cone with half-aperture of $4''$ towards Sgr A*
at $\nu=0.408\,{\rm GHz}$~\cite{Davies},
which we translate in Fig.~\ref{fig:radio} to the $\sigma v$ vs. mass plane.
Let us stress that the  $\sigma v$ plotted in Fig.~\ref{fig:radio} is the \emph{effective} annihilation cross section,
including both Sommerfeld enhancements and boosts due to substructures.
The only way to avoid our constraints would thus be to boost the local anti-matter fluxes by
more than one order of magnitude without affecting the gamma-ray or radio fluxes. Although this theoretical
possibility cannot be ruled out (e.g. Refs.\cite{boost}), it appears to be unlikely
for a realistic distribution of substructures in the Milky Way halo. 
Numerical simulations seem to indicate that the boost factors due to substructure is rather small \cite{Diemand:2008in}. How big the boost factors could be are still under debate and recent simulations \cite{Diemand:2008in} indicate that locally they are at most a factor of a few. A recent study \cite{Afshordi} develops a model that indicates that the local boost could be about a factor of ten. The details of the mechanism giving such large boosts are yet to be presented, however. For more discussion about boost factors, see Ref.~\cite{Bertone:2008xr}.

The two sets of curves give the maximum annihilation cross section
compatible with radio observations of Sgr A* for two different DM
profiles: Einasto and NFW. The shaded region, corresponding to the
range of annihilation cross sections that provide a good fit to the
PAMELA and ATIC data, appears to be in conflict with observations,
unless the DM profile is more shallow than  expected in current
models of structure formation. However, if the DM interpretation of
the PAMELA data was corroborated by additional evidence, then our
result can be interpreted as a hint of the shallowness of the DM
profile.

Profiles steeper than NFW  -- like the $\rho(r) \propto r^{-1.2}$ needed
to explain the WMAP 'Haze' \cite{wmap-haze} -- are ruled out by a rather larger margin.
This confirms the dramatic importance of the multi-wavelength approach to DM studies~\cite{Bertone:2008xr,Bertone:2002ms,radiogc,Regis:2008ij},
especially for DM models tailored to explain anomalies in astrophysical observations.

\paragraph{Acknowledgements.}
LB and JE thank the Swedish Research Council (VR) for support. LB, TB and JE wish to thank IAP, Paris, for hospitality when this work was initiated.


\end{document}